\documentstyle[prd,aps,graphicx]{revtex}
\begin{document}
\draft
\title{Effective Chiral Theory for Radiative Decays of Mesons}
\author{ E.Gedalin\thanks{gedal@bgumail.bgu.ac.il},
 A.Moalem\thanks{moalem@bgumail.bgu.ac.il}
 and L.Razdolskaya\thanks{ljuba@bgumail.bgu.ac.il}}
\address
{ Department of Physics, Ben Gurion University, 84105,
Beer Sheva, Israel}
\maketitle
\begin{abstract}An extended  $U(3)_L\bigotimes U(3)_R$ chiral 
theory which includes pseudoscalar and vector meson nonets as dynamic
degrees of freedom is presented. We combine a hidden symmetry
approach with a general procedure of including the $\eta'$ meson
 into chiral theory. The $U(3)_L\bigotimes U(3)_R$
and the $SU(3)$ symmetries are broken by the mechanism based on quark
mass  matrix. Meson radiative decay widths are parameterized in terms of
a single and real $\eta-\eta'$ mixing angle $\theta_P$, a $U(3)_V$ 
symmetry breaking scale parameter $c_W$, and the radiative decay constants
$F_\pi,~~ F_K, ~~F_\eta, ~~ F_{\eta'}$, for the $\pi , \eta$, K, $\eta '$
mesons, respectively. Taking $F_\pi = 93 MeV$,  a
global fit to decay width data yields
\begin{eqnarray*}
                F_K/F_\pi = 1.16 \pm 0.11~,&~ F_\eta/F_\pi = 1.14\pm 0.04~,&
        ~F_{\eta'}/F_\pi = 1.09 \pm 0.04\\
        c_W = -0.19 \pm0.03~,& ~ \theta_P =-(15.4 \pm 1.8)^o~. &
\end{eqnarray*}
\\


\end{abstract}
\ \\

\bigskip

\bigskip
\section{Introduction}

Meson decays of pseudoscalar and vector mesons have been discussed by
several groups\cite{gilman87,bramon90,bramon97,ball96,bramon99}, using 
phenomenological approaches based on effective field theory.
 In particular, the value of the $\eta$-$\eta '$ mixing angle,
 $\theta_P$, was deduced from the analysis of electromagnetic 
decays of pseudoscalar and vector mesons, $J/\psi$ decays
into a vector and a pseudoscalar meson, and some other transitions.
Gilman and Kauffman\cite{gilman87} assumed SU(3)
symmetry and often the stronger condition of nonet symmetry in order to
relate the SU(3)-octet wave function to that of the SU(3) singlet, and
obtained a value of $\theta_P \cong -20^o$.
Less negative a value was extracted by Bramon and
Scadron\cite{bramon90,bramon97} from a rather similar
analysis which takes into account small departure from the $\omega$-$\phi$
ideal mixing. Somewhat different approach was taken by Ball, Fr\`{e}re and 
Tytgat\cite{ball96} by 
relating vector meson decays to the Quantum Electrodynamics (QED)
triangle anomaly. More recently, Bramon, Escribano and Scadron
\cite{bramon99} have extracted a value $\theta_P = 15.5^o
\pm 1.3^o$ from a rather exhaustive analysis of data including strong
decays of tensor and higher-spin  mesons.

Spontaneously broken chiral symmetry plays a major role in low energy
hadron physics. The Quantum Chromodynamics (QCD)
Lagrangian exhibits an  $SU(3)_L\bigotimes SU(3)_R$ chiral symmetry which 
breaks down spontaneously to  $SU(3)_V$, giving rise to a light
Goldstone boson octet of pseudoscalar mesons. The corresponding
effective field theory (EFT) exhibits the symmetry properties of QCD and
involves both of the pseudoscalar meson octet and vector meson nonet as  
dynamical field variables (see, for example, Ref.\cite{bramon95}).
The axial $U(1)$ symmetry of the QCD Lagrangian is broken by the anomaly. 
Though considerably heavier than the octet states, it is rather well 
accepted that the $\eta '$ meson is the most natural candidate 
for the corresponding pseudoscalar singlet. In this context, we 
shall introduce the $\eta'$ meson also as a dynamical 
field variable. We combine the "hidden symmetry approach"
of Bando et  al.\cite{bando} with a general procedure of including the
$\eta'$ meson into chiral theory\cite{gasser85,leut96,leut97,herera97} to
construct a most general chiral effective Lagrangian with broken 
$U(3)_L\bigotimes U(3)_R$ local symmetry. To this aim we introduce
in section II the whole nonents of pseudoscalar  and vector mesons 
interacting with external electroweak fields. In section III
we apply this approach to study radiative decays for anomalous
processes, like $V^0 \to P^0 \gamma$, $P^0 \to V^0 \gamma$ and
$P^0 \to \gamma \gamma$, with $P^0 = \pi ,\eta ,\eta '$ and $V^0 = 
\rho ,\omega ,\phi$. The numerical values of the radiative decay
constants, $F_i~ ( i=\pi,\eta,K,\eta ')$ and the $\eta$-$\eta '$ 
mixing angle, $\theta_P$ are fixed by fitting to experimental rates of 
these processes. We shall summarize and conclude in section IV. 

\section{The effective Lagrangian}

In order to include the $\eta'$ meson into chiral effective Lagrangian 
one has to extend the $SU(3)_L\bigotimes SU(3)_R$ local symmetry of the
QCD Lagrangian into $U(3)_L\bigotimes U(3)_R$ local symmetry. 
This can be achieved by adding to the QCD Lagrangian (herein denoted 
$L_{QCD}$) a term proportional to the topological charge operator, i.e.,
\begin{equation}
	L = L_{QCD} - 
\Theta(x)\frac{g^2}{16\pi^2}Tr_c\left(G_{\mu\nu}\tilde{G}^{\mu\nu}\right)~,
\label{lqcd}
\end{equation}
where  $\Theta(x)$ represents an auxiliary external field, the
so called QCD vacuum angle.
Here, $\tilde{G}^{\mu\nu} = \epsilon^{\mu\nu\alpha\beta}G_{\alpha\beta}; 
{G}_{\mu\nu} = \partial_\mu G_\nu - \partial_\nu G_\mu +i[G_\mu, G_\nu] 
$, $G_\mu$ being the gluon field, and $Tr_c$ stands for the trace 
over color indices. Obviously, the Lagrangian $L$ of 
Eqn. \ref{lqcd} has $SU(3)_L\bigotimes SU(3)_R$ local symmetry. 
It can be shown \cite{gasser85,leut96,leut97,herera97,herera98}, 
that $L$ would also have $U(3)_L\bigotimes U(3)_R$ local symmetry 
provided $\Theta(x)$ transforms under axial $U(1)$  rotations as, 
\begin{equation}
	\Theta(x) \rightarrow \Theta'(x) = \Theta(x) - 2N_f\alpha~,
\end{equation}
where $N_f$ represents the number of flavors and $\alpha$ the axial
$U(1)$ transformation parameter.
Indeed, 
the term generated by the anomaly in the 
fermion determinant is compensated by the shift in $\Theta(x)$, so that
the overall change in the Lagrangian amounts to a total derivative,
giving rise to the well known anomaly Wess-Zumino term. 
An effective field theory Lagrangian which involves an
integrated form of this anomaly term would also have this same feature.
For more details see Ref. \cite{bramon95}.

We now turn to construct a general chiral effective  Lagrangian with
$U(3)_L\bigotimes U(3)_R$ local symmetry for 
pseudoscalar and vector meson nonets interacting with external
electroweak fields. As a non-linear representation of 
a Goldstone nonet we take \cite{gasser85,leut96},
\begin{equation} 
        U(P,\eta_0+F_0\Theta) = \xi^2(P,\eta_0+F_0\Theta) =
        \exp \left\{i\frac{\sqrt{2}}{F_8}P +
        i\sqrt{\frac{2}{3}}\frac{1}{F_0}(\eta_0 + F_0 \Theta){\bf
1}\right\}~,
        \label{pmfield}
\end{equation}
where $P$ is  the Goldstone pseudoscalar octet,
 \begin{equation}
   P = \left(\begin{array}{ccc}
	\frac{\pi^0}{\sqrt{2}}+\frac{\eta_8}{\sqrt{6}} & \pi^+ &K^+  \\
	\pi^- & - \frac{\pi^0}{\sqrt{2}}+\frac{\eta_8}{\sqrt{6}}  
	&K^0  \\
	K^- & \bar{K}^0 &-\frac{2\eta_8}{\sqrt{6}}
	\end{array}\right)~.
	\label{poctet}
	\end{equation}
and $\eta_0$ the pseudoscalar singlet. The unimodular part of the 
field $U$ contains the octet degrees of freedom while the phase
$detU = \exp {i X} = \exp\left\{i\sqrt{6}(\eta_0/F_0 +
\Theta)\right\}$ describes that of the singlet.
The auxiliary field $\Theta(x)$ ascertains that $detU$ is
invariant under $U(3)_L\bigotimes U(3)_R$ transformations
\cite{gasser85,leut96}. The $U(3)_L\bigotimes U(3)_R$ group does 
not have a dimension-nine irreducible representation; the quantity 
in the curly bracket of Eqn. \ref{pmfield} does not exhibit 
a nonet symmetry so that the octet ($F_8$) and singlet ($F_0$) decay
constants are not necessarily identical. As in Refs.
\cite{gasser88,krause90,bernard95} we define vector 
type $\Gamma _\mu$ and axial-vector type $\Delta_\mu$ covariants
\begin{eqnarray}
	&& \Gamma_\mu = \frac{i}{2}
    \left[\xi^\dagger {\cal D}_\mu \xi - \xi {\cal D}_\mu \xi^\dagger \right] = 
   \frac{i}{2}\left[\xi^\dagger ,\partial_\mu\xi\right]
	  +\frac{1}{2}\left(\xi^\dagger r_\mu \xi + \xi l_\mu\xi^\dagger\right) ~,
	\label{gammac} 
    \\
	&&\Delta_{\mu} =
	 \frac{i}{2}\left(\xi^\dagger{\cal D}_{\mu}\xi + 
	 \xi{\cal D}_{\mu}\xi^\dagger\right)=
     \frac{i}{2}\left\{\xi^\dagger ,\partial_{\mu}\xi\right\}
  +\frac{1}{2}\left(\xi^\dagger r_\mu\xi - \xi l_\mu\xi^\dagger\right) ~,
	\label{deltaf} 
    \end{eqnarray}
 with, 
 \begin{equation}
 	D_\mu \xi = \partial_\mu \xi +i r_\mu \xi - i \xi l_\mu~.
 \end{equation}
Here $r_\mu$ and $l_\mu$ are the relevant external gauge fields of the
standard model; $l_\mu= v_\mu + a_\mu$  and $ \quad l_\mu = v_\mu -
a_\mu$, with $v_\mu$ and  $a_\mu$ being the vector and axial 
vector external electroweak fields, respectively. Electroweak 
interactions are contained in the covariant derivative $D_\mu\xi$.
For pure electromagnetic interactions these fields are related to the
quark charge operator $Q=diag (2/3,-1/3,-1/3)$ and the photon filed
$A_\mu$;~~ $l_\mu = r_\mu = -eQA_\mu$.

  Under $U(3)_L\bigotimes U(3)_R$ the field, Eqn. \ref{pmfield}, 
transforms as,
  \begin{equation}
  	U' = RUL^\dagger, 
  \end{equation}
  with $R\in U(3)_R, \quad L\in U(3)_L$. The vector ($\Gamma _\mu$) and
axial-vector ($\Delta_\mu$) like
quantities  transform, respectively, as a gauge and  matter fields,
i.e.,
  \begin{eqnarray}
  	 & & \Gamma '_\mu = K \Gamma_\mu K^\dagger + iK\partial_\mu 
K^\dagger ~,
  	 \\
  	 & & \Delta' _\mu = K \Delta_\mu K^\dagger~,
  \end{eqnarray}
  where $K(U,R,L)$ is a compensatory field representing an element of 
  conserved vector subgroup $U(3)_V$. 
  
  The dynamical gauge bosons are defined as  a 
  $3 \times 3$ matrix vector field $V_\mu$ which transforms as, 
  \begin{equation}
 V'_\mu =  KV_\mu K^\dagger + \frac{i}{g} K\partial_\mu K^\dagger \ .  
  \end{equation}
Clearly, the vectors $\Gamma_\mu -g V_\mu$ and $\Delta_\mu$ transform  
homogeneously and at lowest order the Lagrangian  can be constructed from  
the traces $Tr\Delta^2_\mu$ ,~~ $Tr(\Gamma_\mu - gV_\mu)^2$, ~~
$Tr\Delta_\mu$,~~ $Tr(\Gamma_\mu - gV_\mu)$ and arbitrary functions 
of the variable $X = \sqrt{2N_f}\eta_0/F_0 +\Theta(x)$, all 
being invariant under $U(3)_L\bigotimes U(3)_R$ transformations.
We may thus conclude that a most general lowest order 
(i.e. with the smallest number of derivatives) effective
chiral Lagrangian can be written in the form \cite{bando},
\begin{equation}
	L = L_A + aL_V - \frac{1}{2}Tr(V_{\mu\nu}V^{\mu\nu})~ ,
\end{equation}
where,
\begin{eqnarray}
	 &  & L_A = W_1(X)Tr(\Delta_\mu\Delta^\mu) 
	 +W_4(X)Tr(\Delta_\mu)Tr(\Delta^\mu) +
	 \nonumber\\
	 && W_5(X)Tr(\Delta_\mu)D_\mu\Theta +
	 W_6(X)D_\mu\Theta D^\mu \Theta~,
	\\
	 &  & L_V =
	 \tilde{W}_1(X)Tr([\Gamma_\mu - gV_\mu][\Gamma^\mu - gV^\mu])+
	 \nonumber\\ 
	 &  & \tilde{W}_4(X)Tr(\Gamma_\mu - gV_\mu)Tr(\Gamma^\mu -
gV^\mu)~,
	 \label{laginv}
\end{eqnarray}
and,
\begin{eqnarray}
	 &  & 	D_\mu \Theta = \partial_\mu \Theta + Tr(r_\mu - l_\mu)~,
	\\
	 &  & V_{\mu\nu} = \partial_\mu V_\nu - \partial_\nu V_\mu -ig 
	 [V_\mu,V_\nu] ~.
\label{vmunu}
\end{eqnarray}
All three terms of the lagrangian $L$ in Eqn. 12 are invariant 
under $U(3)_L\bigotimes U(3)_R$ transformations. 
Though in form this Lagrangian is similar to that of 
Bando et al. \cite{bando}, the expressions for $L_A$ and $L_V$ 
are different. Namely, the inclusion of the
$\eta '$ meson as a dynamical variable involves additional terms with 
$Tr (\Delta^{\mu})$, $D^{\mu} \Theta$ and $Tr(\Gamma_{\mu} - g V_{\mu})$
and coefficient functions $W_i(X)$ which are absent in the $SU(3)$ 
limit. We note though that as in Bando et al. \cite{bando},  
the Lagrangian $L_A + aL_V$ contains, amongst other contributions,      
a vector meson mass term $\sim V_\mu V^\mu$, a vector-photon 
conversion factor $\sim V_\mu A^\mu$ and coupling of both vectors and 
photons to pseudoscalar pairs. The latter can be eliminated by 
choosing $a=2$, a choice which allows incorporating the conventional 
vector-dominance in electromagnetic form-factors of
pseudoscalar mesons\cite{bando}.

The mass degeneracy is removed via the additional of pseudoscalar mass
term. The most general expression of a
(local) $U(3)_L\bigotimes U(3)_R$ symmetry violating term reads 
\cite{gasser85,leut96,herera97},
\begin{equation}
	L_m = -W_0(X) + W_2(X)Tr \chi_+ +iW_3(X) Tr\chi_-~,
\label{maslag}
\end{equation}
with,
\begin{eqnarray}
	 \chi_\pm = 2B_0(\xi{\cal M^\dagger}\xi
	 \pm \xi^\dagger {\cal M} \xi^\dagger)~,
      && \qquad B_0 = m_\pi^2 / (m_u +m_d)~, 
\end{eqnarray}
and ${\cal M} = diag (m_u, m_d, m_s)$ is the quark mass matrix. 
 Parity conservation 
 implies that all $W_i$ and $\tilde{W}_i$ are even functions of the 
variable $X$ except $W_3$ which is odd. The correct normalization 
  of the $U(3)_L\bigotimes U(3)_R$ invariant kinetic term requires that 
  $W_1(0) = F^2_8, \quad W_4(0 )= (F^2_0 - F^2_8)/3$ and $W_2(0) = F^2_8$ 
 to ensure the standard $\chi$PT pion mass term.

  One possible way to incorporate  $SU(3)$ symmetry  breaking is to 
introduce a universal matrix $B$ proportional to $\chi_+$, i.e.,
\begin{equation}
	B = \frac{1}{4B_0(2m + m_s)} \chi_+~.
	\label{breakm}
\end{equation}
For simplicity we take the exact isospin symmetry limit $m_u = m_d = m$.
Symmetry breaking terms to be added to $L_A$ and $L_V$ can be constructed
either as conserving,  or alternatively, as non-conserving the quadratic
form of the Golstone meson kinetic terms. In what follows we develop 
the former alternative by including terms which break the octet 
symmetry only. The latter procedure is worked out in the Appendix. Let,
\begin{equation}
	U_8 = \xi^2_8 = \exp(i \frac{\sqrt{2}}{F_8}P)
	\label{u8}
\end{equation}
be the pure octet matrix and let,
\begin{equation}
	 \bar{\Delta}_\mu = 
	 \frac{i}{2}\left\{\xi^\dagger_8 ,\partial_{\mu}\xi_8\right\}
 +\frac{1}{2}\left(\xi^\dagger_8 r_\mu\xi_8 - 
	  \xi_8 l_\mu\xi^\dagger_8\right) ~,
	\label{bdelta} 
\end{equation}
be the  octet covariant. Then general $SU(3)$ symmetry breaking 
Lagrangians $\bar{L}_A$ and $\bar{L}_V$ would be,
\begin{eqnarray}
	 &  &\bar{L}_A =
	 \nonumber\\
	  && W_1(X)\left( c_A Tr (\{B, \bar{\Delta}_\mu\} 
	  \bar{\Delta}^\mu) + 
	 d_A Tr (B \bar{\Delta}_\mu B\bar{\Delta}^\mu)\right)+
	 \nonumber \\
	 && W_4(X)
	 d_A Tr (B \bar{\Delta}_\mu) Tr (B\bar{\Delta}^\mu)~, 
	\label{labreak} \\
	 &  & \bar{L}_V =
	 \bar{W}_1(X)\left(c_V Tr (B[\Gamma_\mu -gV_\mu] 
	[\Gamma^\mu -gV^\mu])\right. +
	\nonumber\\
	&& \left. d_V Tr (B[\Gamma_\mu-gV_\mu] B [\Gamma^\mu -gV^\mu])\right) +
	 \nonumber\\
	 &&\tilde{W}_4(X)\left(c_VTr(\Gamma_\mu - gV_\mu)Tr(B[\Gamma^\mu -
           gV^\mu])\right.
	 \nonumber\\
	&& +\left. d_VTr(B[\Gamma_\mu - gV_\mu])Tr(B[\Gamma^\mu -
         gV^\mu])\right)~.
	\label{lvbreak}
\end{eqnarray}
where $c_A, c_V, d_A, d_V$ are arbitrary constants. We stress that
$\bar{L}_A$ and $\bar{L}_V$ differ from those of Bramon et
al.\cite{bramon95} and Bando et al. \cite{bando}. First, like our
symmetric $L_A$ and $L_V$ the asymmetric $\bar{L}_A$ and 
$\bar{L}_V$ parts involve additional terms which are
absent in the $SU(3)$ limit. Secondly, the terms proportional to $d_A$
and $d_V$ were included by Bando et al. \cite {bando} but with $d_i =
c_i^2$. Thirdly, our symmetry breaking matrix B is not constant as in
ref. \cite{bando} though similar (but not identical) to that of Bramon et
al.\cite{bramon95}.

The full Lagrangian may now be written in the form,
\begin{equation}
	L = L_A + \bar{L}_A + a(L_V + \bar{L}_V) + L_m + L_{WZW} - 
	 \frac{1}{2}Tr(V_{\mu\nu}V^{\mu\nu}) ~,
	\label{elag}
\end{equation}
where we included the well known Wess-Zumino-Witten term $L_{WZW}$. 
This corresponds to the action defined as \cite{witten83,callan84}
\begin{eqnarray}  
	 &  & S_{WZW} = -\frac{i}{80\pi^2}\int_{M^2} d^5x 
	 \epsilon^{ijklm}Tr(\Sigma^L_i\Sigma^L_j\Sigma^L_k\Sigma^L_l\Sigma^L_m) 
	\nonumber\\
	 &  & -\frac{i}{16\pi^2}\int d^4x \epsilon^{\mu\nu\alpha\beta}
	 \left[W(U,l,r)_{\mu\nu\alpha\beta} - 
	 W(1,l,r)_{\mu\nu\alpha\beta}\right]~,
\end{eqnarray}
with,
\begin{eqnarray}
	 &  & W(U,l,r)_{\mu\nu\alpha\beta} = 
Tr[U l_\mu l_\nu l_\alpha l_\beta + 
	 \frac{1}{4}U l_\mu U^\dagger r_\nu U l_\alpha U^\dagger r_\beta
		\nonumber\\
	 &  & + iU\partial_\mu l_\nu l_\alpha U^\dagger r_\beta
	      + iU\partial_\mu r_\nu l_\alpha U^\dagger r_\beta
	      - i\Sigma^L_\mu l_\nu U^\dagger r_\alpha U l_\beta 
		\nonumber\\
	 &  & + \Sigma^L_\mu U^\dagger \partial_\nu r_\alpha U l_\beta 
	      - \Sigma^L_\mu \Sigma^L_\nu U^\dagger  r_\alpha U l_\beta 
	      + \Sigma^L_\mu l_\nu\partial_\alpha l_\beta 
	      + \Sigma^L_\mu \partial_\nu l_\alpha l_\beta 
		\nonumber\\
	 &  & - i\Sigma^L_\mu l_\nu l_\alpha l_\beta 
	 + \frac{1}{2} \Sigma^L_\mu l_\nu \Sigma^L_\alpha l_\beta 
	 - i\Sigma^L_\mu \Sigma^L_\nu \Sigma^L_\alpha l_\beta - 
	 (L \leftrightarrow  R)]~,
	 \label{wzw}
\end{eqnarray}
where $\Sigma^L_\mu = U^\dagger \partial_\mu U$,
      $\Sigma^R_\mu = U\partial_\mu U^\dagger$, 
and $(L \leftrightarrow  R)$ stands for a similar expression with 
$U$, $l$ and $\Sigma$ interchanged according to,
\begin{equation}
	 (U \leftrightarrow  U^\dagger),\qquad (l \leftrightarrow  r),
	 \qquad (\Sigma^L_\mu  \leftrightarrow \Sigma^R_\mu  )~.
\end{equation}
Note that this expression involves Lagrangian terms up to fifth
chiral order, only. Other terms which accounts for the regularization
of the one loop contributions are listed in 
Refs.\cite{gasser85,bijnens90,bijnens901}.

 For the purpose of treating radiative decays, one can safely neglect
the auxiliary field $\Theta(x)$ and the quantities $W_i$ and $\bar{W}_i$ 
 become functions of $\eta_0$ only. To lowest order their expansions read,  
 \begin{eqnarray}
 	 &  & W_0 = const + F^4_8 w_0 \frac{\eta^2_0}{F^2_0} + \ldots~,
 	\label{wco} \\
 	 &  & W_1 = F^2_8(1 + w_1 \frac{\eta^2_0}{F^2_0} + \ldots)~,
 	\label{wc1} \\
 	 &  &  W_2 = \frac{F^2_8}{4}(1 + w_2 \frac{\eta^2_0}{F^2_0}+ \ldots)~,
 	\label{wc2} \\
 	 &  &  W_3 = \frac{ F^2_8}{2}(w_3 \frac{\eta_0}{F_0} + \ldots)~,
 	\label{wc3} \\
 	 &  &  W_4 = \frac{F^2_0 - F^2_8}{3}(1 + w_4 \frac{\eta^2_0}{F^2_0}
 	  + \ldots)~,
 	\label{wc4} \\
 	 &  & \tilde{W}_1 = F^2_8(1 + \tilde{w}_1\frac{\eta^2_0} {F^2_0}+ 
 	 \ldots)~,
 	 \\label{wwc1}¥ \\
 	 &&\tilde{W}_4 = F^2_8(\tilde{w}_4 + \tilde{w}'_4\frac{\eta^2_0}{F^2_0}+ 
 	 \ldots~).
 	\label{wwc4} 
 \end{eqnarray}
 By substituting these expressions into Eqn. \ref{elag}, the kinetic terms
of the pseudoscalar mesons is,
  \begin{eqnarray}
  	 &  & L_{kin} = \frac{1}{2}\left(1 + c_A \frac{2m}{2m+m_s}
  	 + d_A \frac{m^2}{(2m+m_s)^2}\right)(\partial_\mu \vec{\pi})^2+
  \nonumber\\	
  	 &  & \frac{1}{2}\left(1 + c_A \frac{m+m_s}{2m+m_s}
  	 + d_A \frac{mm_s}{(2m+m_s)^2}\right)\sum_{i}(\partial_\mu K_i)^2+
  	\nonumber\\
  	 &  & \frac{1}{2}\left(1 + c_A\frac{1}{3} \frac{m+2m_s}{2m+m_s}
  	 + d_A \frac{1}{3}\frac{m^2+2m^2_s}{(2m+m_s)^2}\right)
  	 (\partial_\mu \eta_8)^2+
  	 \frac{1}{2}(\partial_\mu \eta_0)^2~.
  	\label{lkin}
  \end{eqnarray}
To restore the standard normalization of the kinetic term we rescale the 
pseudoscalar fields according to,
\begin{equation}
	\pi\Rightarrow z_\pi\pi,\qquad K\Rightarrow z_s K,\qquad \eta_8 
	\Rightarrow z_8\eta_8~,
	\label{rescal}
\end{equation}
with,
\begin{eqnarray}
	 &  & z_\pi \equiv \frac {F_8}{F_\pi} 
              = \frac{1}{\sqrt{1 + c_A \frac{2m}{2m+m_s}
  	                  + d_A \frac{m^2}{(2m+m_s)^2}}}~,
	\label{zpi} \\
	 &  & z_s \equiv \frac {F_8}{F_K} 
              = \frac{1}{\sqrt{1 + c_A \frac{m+m_s}{2m+m_s}
  	                    + d_A \frac{mm_s}{(2m+m_s)^2}}}~,
	\label{zs} \\
	 &  & z_8   \equiv \frac {F_8}{F_\eta}
              = \frac{1}{\sqrt{1 + c_A\frac{1}{3} \frac{m+2m_s}{2m+m_s}
  	 + d_A \frac{1}{3}\frac{m^2+2m^2_s}{(2m+m_s)^2}}}~.
	\label{zet}
\end{eqnarray}
After some algebraic manipulations the octet field matrix can be written
in the form,
 \begin{equation}
   P = \left(\begin{array}{ccc}
	\frac{\pi^0}{\sqrt{2}}+\bar{z}_8\frac{\eta_8}{\sqrt{6}} 
	& \pi^+ &\bar{z}_sK^+  \\
	\pi^- & - \frac{\pi^0}{\sqrt{2}}+\bar{z}_8\frac{\eta_8}{\sqrt{6}}  
	&\bar{z}_sK^0  \\
	\bar{z}_sK^- & \bar{z}_s\bar{K}^0 &-\bar{z}_8\frac{2\eta_8}{\sqrt{6}}
	\end{array}\right)~,
	\label{poctet}
	\end{equation}
	where,
	\begin{equation}
		\bar{z}_8 = z_8/z_\pi,  \qquad \bar{z}_s  = z_s/z_\pi ~. 
	\end{equation}
In addition to the usual quadratic terms $\eta^2_8$ and
$\eta^2_0$, the quantity $W_2(X)Tr\chi_+ +iW_3(X)Tr\chi_-$ in the mass
term, Eqn. \ref{maslag} gives rise to a mixing term $\sim \eta_8\eta_0$ 
which violates the orthogonality of the  $\eta_8$ and $\eta_0$ states.
The mass matrix is diagonalized via the usual unitary  transformation
\cite{pdg98}
 \begin{eqnarray}
	 &  & \eta_8 = \eta \cos \theta_P + \eta'\sin \theta_P~, 
 \\
	 &  & \eta_0 = - \eta \sin  \theta_P + \eta' \cos \theta_P~, 
\end{eqnarray}
where $ \theta_P$ is  the $\eta$-$\eta '$ mixing angle.

In terms of the physical fields $\eta$ and $\eta'$, the nonlinear
representation of the pseudoscalar particles can now be written as,
 \begin{equation}
 	U = \exp{i\frac{\sqrt{2}}{F_\pi}{\cal P}}~,
 	\label{ufield}
 \end{equation}
 where $\cal P$ stands for the pseudoscalar nonet matrix,
  \begin{equation}
{\cal P} = \left(\begin{array}{ccc}
		\frac{\pi^0}{\sqrt{2}}+\frac{1}{\sqrt{6}}(X_\eta \eta +
		X_{\eta'}\eta') & \pi^+ &\bar{z}_sK^+  \\
		\pi^- & - \frac{\pi^0}{\sqrt{2}}+ \frac{1}
		{\sqrt{6}}(X_\eta \eta +X_{\eta'}\eta')  &\bar{z}_sK^0  \\
		\bar{z}_sK^- & \bar{z}_s\bar{K}^0 &\frac{1}{\sqrt{6}}
		(Y_\eta \eta + Y_{\eta'}\eta')
	\end{array}\right)~,
	\label{pnonet}
	\end{equation}
	with,
	\begin{eqnarray}
 X_\eta = \cos\theta_P(\bar{z}_8 - \sqrt{2}\bar{r}\tan\theta_P)~, &
 \quad &X_{\eta'} = \cos\theta_P(\bar{z}_8\tan\theta_P +\sqrt{2}\bar{r})~,
    \nonumber	 \\
	    Y_\eta = \cos\theta_P(-2\bar{z}_8 -
\sqrt{2}\bar{r}\tan\theta_P)~, 
 &\quad &Y_{\eta'} = \cos\theta_P(-2\bar{z}_8\tan\theta_P +
\sqrt{2}\bar{r})~,
	    \label{xy} 
	\end{eqnarray}
and $\bar{r} = F_8 / F_0~$.

Similarly, the vector nonet with ideal mixing has the form\cite{pdg98},
	\begin{equation}
     V = \left(\begin{array}{ccc}
		\frac{\rho^0}{\sqrt{2}}+\frac{\omega}{\sqrt{2}} & \rho^+ & K^{*+}  \\
		\rho^- & - \frac{\rho^0}{\sqrt{2}}+\frac{\omega}{\sqrt{2}}  & K^{*0}  \\
		K^{*-} & \bar{K}^{*0} &\phi
	\end{array}\right)~.
	\label{vnonet}
	\end{equation}
In order to account for a strange (non strange) admixture in
$\omega (\phi)$ we substitute  	
\begin{eqnarray}
	& &\omega \rightarrow \omega - \epsilon' \phi ~, \\
	& &\phi \rightarrow \phi +\epsilon' \omega  ~.
\end{eqnarray}
   
\bigskip
\section{ Radiative Decay Widths}

We now turn  to calculate radiative decay widths for 
$P^0 \rightarrow \gamma \gamma$, $V^0 \rightarrow P^0 \gamma$ and 
$P^0 \rightarrow V^0 \gamma$, with $P^0 = \pi ,\eta ,\eta ' , K$
and $V^0 = \rho ,\omega ,\phi , K^*$ using
the formalism outlined above. We generalize  the treatment 
of Ref.\cite{bramon95} by incorporating "indirect" symmetry breaking
 effects via pseudoscalar and vector nonet matrices Eqns. \ref{pnonet}
and \ref{vnonet} and "direct" symmetry breaking terms ( such as  
$\bar{L}_A$ and $\bar{L_V}$). The Lagrangian is factorized in the form,
\begin{eqnarray}
         &  &  L_{P\gamma\gamma} =  L^{(s)}_{P\gamma\gamma} +
          c_W  L^{(b)}_{P\gamma\gamma},
         \\
         &  &  L_{VP\gamma} =  L^{(s)}_{VP\gamma} +
          c_W L^{(b)}_{VP\gamma}~,
         \label{dlag}
\end{eqnarray}
where $L^{(s)}$ and $L^{(b)}$ are generic for  indirect and
direct symmetry breaking contributions, and $c_W$ is a symmetry
breaking parameter. In order to write these terms explicitly, we
consider first the indirect anomalous Lagrangian\cite{bijnens90},
\begin{equation}
	L^{(s)}_{anomalous} =    L^{(0)}_{VVP} +
	 L_{WZW}(P\gamma\gamma)~.
	\label{dlag0}
\end{equation}
From the four covariants $\Delta_\mu$, $\Gamma_\mu-gV_\mu$, 
$V_{\mu \nu}$ and $  \Gamma_{\mu\nu} = \partial_\mu \Gamma_\nu -
\partial_\nu \Gamma_\mu - i[\Gamma_\mu, \Gamma_\nu]$, the Lagrangian
$L^{(0)}_{VVP}$ can have at most six terms,
\begin{eqnarray}
	 &&  L^{(0)}_{VVP} = g_1\epsilon^{\mu\nu\alpha\beta}
	 Tr(V_{\mu\nu} 
	 [V_\alpha -\frac{1}{g}\Gamma_\alpha]\Delta_\beta) +
	 g_2\epsilon^{\mu\nu\alpha\beta}Tr(\Gamma_{\mu\nu} 
	 [V_\alpha -\frac{1}{g}\Gamma_\alpha]\Delta_\beta) +
	 \nonumber \\
	 && g_3\epsilon^{\mu\nu\alpha\beta}Tr(V_{\mu\nu})
	 Tr([V_\alpha -\frac{1}{g}\Gamma_\alpha]\Delta_\beta) +
	  g_4 \epsilon^{\mu\nu\alpha\beta}Tr(\Gamma_{\mu\nu})
	 Tr([V_\alpha -\frac{1}{g}\Gamma_\alpha]\Delta_\beta) +
	 \nonumber \\
	 && g_5\epsilon^{\mu\nu\alpha\beta}Tr(V_{\mu\nu})
	 Tr(V_\alpha -\frac{1}{g}\Gamma_\alpha )Tr(\Delta_\beta) +
	 \nonumber \\
	 && g_6\epsilon^{\mu\nu\alpha\beta}Tr(\Gamma_{\mu\nu})
	 Tr(V_\alpha -\frac{1}{g}\Gamma_\alpha )Tr(\Delta_\beta)~,
	\label{vvplag}
\end{eqnarray}
where $g_i,~~~i=1,...6$
are arbitrary functions of the variable X.
We recall that $\Gamma_\mu$ involves a term proportional 
to the photon field $A_\mu$. By rearranging contributions to
$VP\gamma$ and $P\gamma\gamma$ interaction terms we obtain,
\begin{eqnarray}
	 &  & L^{(s)}_{VP\gamma}= 
	 g_V \frac {e}{F_\pi}\epsilon^{\mu\nu\alpha\beta}\partial_\mu 
	 A_\nu Tr( Q\{\partial_\alpha V_\beta, {\cal P}\})~, 
	\label{vpgamma}\\
	 &  &  L^{(s)}_{P\gamma\gamma} =
	 g_P\frac{e^2}{2F_\pi}\epsilon^{\mu\nu\alpha\beta}\partial_\mu 
	 A_\nu\partial_\alpha A_\beta Tr(\left\{ Q^2,{\cal P}\right\})~.
	\label{pgamma} 
\end{eqnarray}
For convenience we have introduced coupling constants $g_V$ and $g_P$
which incorporate all relevant contributions to 
 $ L^{(s)}_{VP\gamma}$ and $ L^{(s)}_{P\gamma\gamma}$.
It is now rather easy to obtain  the direct symmetry breaking terms 
by introducing the quantity B, Eqn. \ref{breakm}, as described in the
previous section, i.e.,
\begin{eqnarray}
     &  & L^{(b)}_{VP\gamma} = 
     g_V \frac {e}{F_\pi}\epsilon^{\mu\nu\alpha\beta}\partial_\mu 
	 A_\nu Tr( Q \{B,\{\partial_\alpha {V}_\beta ,{\cal P}\}\})~,
	\label{vpgammab} \\
     &  &  L^{(b)}_{P\gamma\gamma}= 
	 g_P\frac{e^2}{2F_\pi}\epsilon^{\mu\nu\alpha\beta}\partial_\mu 
	 A_\nu\partial_\alpha A_\beta Tr(\left\{ Q^2,\{B,{\cal P}\}\right\})~.
	\label{ptwogammab}
\end{eqnarray}

\subsection{The $V\to P \gamma$ and $P \to V \gamma$ Processes}
The relevant vertices are,
 \begin{equation}
 	  V(VP\gamma) = 
	- i g_V\frac{e}{F_\pi}w(VP)\epsilon^{\mu\nu\alpha\beta}k_\mu 
	 e^{(\gamma)}_\nu p_\alpha e^{(V)}_\beta ~,
 	\label{vpgamma} 
 \end{equation}
 where $ e^{(V)}_\nu$ ($p$) and $ e^{(\gamma)}_\nu$ ($k$) are 
 the polarization (four-momentum) of the vector meson and final photon,
 respectively. With the quark mass ratios 
 advocated by Weinberg\cite{wein77} $m_u:m_d:m_s= 0.55:1.0:20.3$, 
 the ratio $m:m_s=(m_u+m_d):2m_s= 0.038$ is rather small and terms 
 proportional to $c_W m/(2m+m_s)$ can be neglected.\footnote{The
Weinberg's ratios give apparently the lowest limit for $m_s/m$. 
The current algebra prediction is $m_s/m = (2m^2_K -m^2_\pi)/m^2_\pi=25.6$ 
while recent estimations\cite{leut97} give the value $m_s/m \approx 26.6$.}.
Then for the ${\cal P}$- matrix of  Eqns. \ref{pnonet}, \ref{xy} one
obtains,
\begin{eqnarray}
   w(\rho \pi) = \frac{1}{3}~,& \quad w(\rho \eta) =
\frac{1}{\sqrt{3}} X_\eta ~, &\quad w(\rho \eta') = \frac{1}{\sqrt{3}}
X_{\eta'}~,
\label{rhop}\\
   w(\omega \pi) = 1~,& \quad w(\omega \eta) = - \frac{1}{3\sqrt{3}}
X_\eta~,& \quad w(\omega \eta') = \frac{1}{3\sqrt{3}}  X_{\eta'}~,
\label{omegap}\\
   w(\phi \pi) = \epsilon'~,&\qquad  w(\phi \eta) = -
\frac{\sqrt{2}}{3\sqrt{3}} 
 	Y_\eta (1+ c_W \frac{m_s}{2m+m_s})~,&
\nonumber\\
  & \qquad w(\phi \eta') = - \frac{\sqrt{2}}{3\sqrt{3}}
 	 Y_{\eta'}(1 + c_W\frac{m_s}{2m+m_s})~,&
\label{phip}
\end{eqnarray}
\begin{eqnarray}
  && w(K^{*0}K^0) = w(\bar{K}^{*0}\bar{K}^0)
 	  = -\frac{2}{3}\bar{z}_s (1+\frac{1}{2}c_W \frac{m_s}{2m+m_s})~,  
 \label{kstk0}\\
 &&  w(K^{*+}K^+) = w(K^{*-}K^-) 
 	  = \frac{1}{3}\bar{z}_s (1-c_W \frac{m_s}{2m+m_s})~. 
\label{kstpkp} 	 
\end{eqnarray}

In terms of these vertices the decay widths of
 $V \rightarrow P \gamma$ and $P \rightarrow V \gamma$ are,
\begin{eqnarray}
 	 &  & \Gamma (VP\gamma) = G\frac{(m^2_V - m^2_P)^3}{m^3_V F^2_8}
 	|w(VP)|^2~,
 \label{gvp}	 \\
     &  & \Gamma (P V\gamma) =3 G\frac{(m^2_{P} - m^2_V)^3}
     {m^3_{P} F^2_8}|w(V P)|^2~,
     \label{pvg}	
 \end{eqnarray}
 with, 
 \begin{equation}
 	G = \frac{e^2}{4\pi}\frac{g^2_V}{24}~.
 \end{equation}

\subsection{The $P\rightarrow \gamma\gamma$ decays}
The relevant vertices are,
 \begin{equation}
 	  V(P\gamma\gamma) = 
	- 2i g_P\frac{e^2}{F_8}\bar{w}(P)\epsilon^{\mu\nu\alpha\beta}k_{1\mu} 
	 e^{(\gamma)}_\nu k_{2\alpha} e^{(\gamma)}_\beta ~,
 \end{equation}
 where $ e^{(\gamma)}_\nu$ and $ e^{(\gamma)}_\alpha$ are 
 the polarizations of the final photons , $k_1$ and $k_2$ are their 
 corresponding four-momenta. For the $\pi , \eta $ and $\eta '$ one has, 
 \begin{eqnarray}
   & &	\bar{w}(\pi) = \frac{3}{\sqrt{2}}~,
 \label{wpi}¥	 \\
   & & \bar{w}(\eta) = 
   \nonumber\\
   &&\frac{3\cos \theta_P}{\sqrt{6}}\left[\bar{z}_8(1-\frac{4}{3}c_W
   \frac{m_s}{2m+m_s})
    -2\sqrt{2}(1+\frac{1}{3}c_W \frac{m_s}{2m+m_s})\bar{r}\tan \theta_P
\right]~, 
\label{weta}      \\	 
   & & \bar{w}(\eta') = 
   \nonumber \\
   &&\frac{3\cos\theta_P}{\sqrt{6}}\left[\bar{z}_8(1-\frac{4}{3}c_W
   \frac{m_s}{2m+m_s})\tan\theta_P
    +2\sqrt{2}(1+\frac{1}{3}c_W \frac{m_s}{2m+m_s}) \bar{r}\right]~.
   \label{weta'}¥
  \end{eqnarray}
With these vertices the decay  rate is given by,
 \begin{equation}
 	  \Gamma (P\gamma\gamma) = \bar{G}\frac{ m^3_P}{ F^2_8}|\bar{w}(P)|^2~,
 	  \label{p2gamma}¥
 \end{equation}
 with,
 \begin{equation}
 	\bar{G} = \frac{\pi}{2}\left(\frac{e^2}{4\pi}\right)^2\left(\frac{g_P }
 	{9}\right)^2~.
 \end{equation}

\subsection{Numerical Analysis and Results} 

The decay width of the anomalous processes mentioned above are
described in terms of coupling constants ($g_V ,g_P$), pseudoscalar 
singlet-octet mixing angle ($\theta_P$), relative radiative decay
constants (${\bar r}, {\bar z}_s, {\bar z}$) and direct $U(3)_V$ symmetry
breaking scale ($c_W$).  The numerical values of these parameters can be
fixed from experimental decay rates. 
The value of $g_V$ is determined from the $\omega \rightarrow \pi \gamma$
decay, $\Gamma(\omega\pi\gamma) =  G (m^2_\omega - m^2_\pi)^3/
        (m^3_\omega F^2_8) = (716 \pm 43) KeV$, to have,
\begin{eqnarray}
 G=(1.44\pm0.04)\cdot10^{-5}~, &\qquad &
        g_V = 0.22 \pm 0.006~.
        \label{gvec}
   \end{eqnarray}
From the $\rho \rightarrow \pi \gamma$ decay,  $\Gamma(\rho\pi\gamma) = 76\pm 10
KeV$ one obtains practically identical value for $g_V$. 
Similarly, the decay $\pi \rightarrow \gamma \gamma$ can serve to fix $g_P$.
From the experimental decay width, 
       $ \Gamma(\pi^0\gamma\gamma) = 9 \bar{G} m^3_\pi/2F^2_8 
        = (7.8 \pm 0.55) eV$, one obtains,
\begin{eqnarray}
\bar{G}=(4.9\pm0.07)\cdot10^{-8}, &\qquad &
        g_P = 0.073 \pm 0.001~.
        \label{gps}
   \end{eqnarray}
The value of the symmetry breaking scale $c_W$ can be fixed from the ratio, 
 \begin{eqnarray}
         &&\frac{\Gamma (K^{*0}K^0\gamma)}{\Gamma (K^{*+}K^+\gamma)} =
         4\left[\frac{1+\frac{1}{2}c_W }{1-c_W}\right]^2 =
         \frac{(117\pm10)KeV}{(50\pm 5)KeV} = 2.34\pm 0.43~,
        \label{cwpar}
 \end{eqnarray}
which yields $c_W = -0.19 \pm 0.04$. From equating 
$\Gamma(K^{*0}K^0\gamma)$ to its experimental value  one 
 finds $\bar{z}_s = 0.86 \pm 0.08$ a value corresponding to 
$F_K = (1.16 \pm 0.11)F_\pi$. In fact, from the ratio,
 \begin{eqnarray}
         &  & \frac{\Gamma(\phi\pi^0\gamma)}{\Gamma(\omega\pi^0\gamma)} =
        \epsilon'^2 \left[\frac{(m^2_\phi - m^2_\pi)m_\omega}
         {(m^2_\omega - m^2_\pi)m_\phi}\right]^3 =
        \frac{(5.8 \pm 0.6)KeV}{(716\pm 43) KeV} = 0.008 \pm 0.001
\end{eqnarray}
one obtains, $ |w(\phi \pi^0)| = 0.059 \pm 0.005$, a value identical to
that quoted previously by Bramon et al. \cite{bramon95} from using the 
vector meson dominance.

  The remaining parameters $\bar{z}$, $\bar{r}$ and  $\phi_P$ 
  can now be calculated using Eqns. 
\ref{gvp}, \ref{pvg}, \ref{p2gamma} and the experimental decay widths
   $\Gamma(\eta\gamma\gamma)$, $\Gamma(\phi\eta\gamma)$
 	  and $\Gamma(\eta'\gamma\gamma)$. From the ratios 
$\Gamma(\eta\gamma\gamma)/ \Gamma(\phi\eta\gamma)$, 
   $\Gamma(\eta\gamma\gamma)/\Gamma(\pi^0\gamma\gamma)$, and
   $\Gamma(\eta'\gamma\gamma)/\Gamma(\eta\gamma\gamma)$, one obtains,
   \begin{eqnarray}
   	  \bar{z} = 0.92\pm0.06~, &\quad \bar{r} = 0.97 \pm0.06~,
   	& \quad\theta_P = -(15 \pm 2.4)^o~.
   	\label{¥}
   \end{eqnarray}
These values (hereafter we refer to as solution I) are listed 
in Table \ref{fit}. Rather similar values (solution II of 
Table \ref{fit}) one deduced from a global fit of the
data listed in Table \ref{vpgpred}. Predictions of decay widths as
obtained with the solutions I and II of Table \ref{fit}
 are summarized in Table \ref{vpgpred}. 

From Eqns. ~\ref{rhop}-\ref{phip} and Eqns.~ \ref{wpi} - \ref{weta'}  
one can see that for fixed $c_W$ the 
width ratios $\Gamma(\eta'\gamma\gamma)/ \Gamma(\eta\gamma\gamma)$,
$\Gamma(\phi\eta'\gamma)/ \Gamma(\phi\eta\gamma)$,
$\Gamma(\eta'\omega\gamma)/ \Gamma(\omega\eta\gamma)$ and 
$\Gamma(\eta'\rho\gamma)/ \Gamma(\rho\eta\gamma)$ depend on 
$\tan\theta_P$ and $r/z$. Precision measurements of these 
ratios would be very useful to obtain more accurate values for 
$\theta_P$ and $r/z$. Perhaps even more attractive quantities are the
ratios $\Gamma(\eta'\gamma\gamma)/ \Gamma(\eta\gamma\gamma)$,
$\Gamma(\phi\eta'\gamma)/ \Gamma(\phi\eta\gamma)$ which can now be 
determined at DA$\Phi$NE with high precision.

Our set of parameters agree with the results of Bramon et
al.\cite{bramon95} and Venugopal et al. \cite{veno98} except for
the mixing angle which differs significantly from  Bramon et
al.\cite{bramon95} and  Venugopal et al. \cite{veno98} but agrees with
more recent analysis of Bramon et al.\cite{bramon97} and Escribano et
al.\cite{escrib99}.

How significant are the departure of these parameters from their values 
in the limit exact $U(3)_L\bigotimes U(3)_R$ symmetry?
Clearly, the exact $SU(3)$ limit, i.e., $c_W =0, F_\pi = F_K = F_8$ is
inconsistent with the ratio $\Gamma (K^{*0}K^{0}\gamma)/
\Gamma (K^{*+}K^{+}\gamma)$; a value of $c_W = 0$ predicts a ratio equals
4 as opposed to the experimental value of $2.34 \pm 0.43$. Furthermore,
with $c_W = - 0.19$ and with $F_K = F_\pi$ one obtains $\Gamma
(K^{*0}K^{0}\gamma)$ about 35$\%$ higher than experimental value and far 
beyond the measurement
accuracy. We may thus conclude that data requires $SU(3)$ symmetry to be
broken directly ($c_W \neq 0$) and indirectly ($F_K \neq F_\pi$). If 
either direct or indirect symmetry breaking is not included, the quality 
of the fit deteriorates significantly. The
value of the mixing angle $\theta_P$ is rather sensitive to direct
symmetry breaking. At the limit $F_0 = F_8$ the mixing angle varies from 
$\theta_P = -23^o $ at $c_W = 0$ to $\theta_P = -16^o $ at $c_W =
-0.19$. The mixing angle is less sensitive to indirect symmetry
breaking. Indeed, a global fit which neglects indirect symmetry
breaking,i.e., with $F_K = F_\pi = F_8 = F_0$
gives $c_W = -0.22$ and $\theta_P = -16.2^o$. Also, a global fit
which assumes broken SU(3) symmetry but with nonet symmetry , i.e., 
with $F_0 = F_8 =F_K = F$ but $F \neq F_\pi$ yields  $F = 1.1 F_\pi$,
$c_w = -0.20$ and $\theta_P = -14.6^o$, rather close to the values of 
solution II. Upon concluding we stress that confidence criterion 
favorables our solution II, i.e., with direct and indirect  symmetry 
breaking. 

\section{Summary and Discussion}¥

In this paper, using the hidden symmetry approach of Bando 
et al.\cite{bando} combined with general procedure of including the 
$\eta'$ meson into $\chi$PT\cite{gasser85,leut96,leut97,herera97} we
 have constructed an effective  Lagrangian which incorporates 
pseudoscalar and vector meson nonets as dynamical degrees of freedom 
 interacting with external electroweak fields. At lowest order 
the  Lagrangian $L$ is a linear combination of three parts $L_A$, $L_V$ 
and a vector nonet "kinetic" term $\frac{1}{2}Tr(V_{\mu\nu}V^{\mu\nu})$, 
 all of which possessing a $U(3)_L\bigotimes U(3)_R$ and a local (hidden)
 $U(3)_V$ symmetry. The $L_A$ and $L_V$ parts involve the pseudoscalar
and vector meson fields and their interactions   
with external electroweak fields, respectively.
Though in form this division of the Lagrangian is identical to that
of Bando et al.\cite{bando}, the expressions for $L_A$ and $L_V$ are
different, as they include the $\eta '$ meson as a dynamical variable
also. The symmetry breaking effects are included via the pseudoscalar 
meson mass term as well as direct symmetry breaking terms ${\bar
L}_A$ and ${\bar L}_V$ in a fashion similar to that proposed by Bramon 
et al. \cite{bramon95}. These terms are constructed by introducing 
a universal matrix $B$ which is proportional to  pseudoscalar meson 
mass matrix into our general expressions for $L_A$ and $L_V$. 
The symmetry breaking leads to the mass splitting 
for the pseudoscalar and vector meson nonets, $\eta-\eta'$ and 
$\omega-\phi$ mixing effects, $F_\pi\neq F_K\neq F_\eta \neq F_{\eta'}$ etc.
We may thus conclude that the Lagrangian of Eqn.\ref{elag} provides a basis
for an effective perturbative chiral theory capable to describe 
interacting pseudoscalar and vector mesons. To demonstrate that, 
we have considered anomalous radiative decay processes within our 
 approach. Namely, the decay widths of anomalous processes are
calculated by taking into account indirect as well as direct symmetry 
breaking effects. The widths were parameterized in terms of five 
parameters, including a symmetry breaking scale $c_W$,  pseudoscalar 
meson weak decay constants $F_K,~~~F_\eta,~~~F_{\eta'}$ and
the $\eta-\eta'$ meson mixing angle $\theta_P$. Our analysis show that
the value of the mixing angle $\theta_P$ is rather sensitive to the 
presence of a direct symmetry breaking. The best solution was obtained
with $c_W = -0.19$ suggesting a value 
$\theta_P \approx -(15.4 \pm 1.8)^o$. This agrees with the value extracted
by Bramon et al. \cite{bramon99} from rather exhaustive analysis of data.
Our analysis provides evidence for a broken $U(3)$ symmetry with $F_0
\neq F_8$ and $F_K \neq F_\eta \neq F_\pi$. 

\bigskip
{\bf Acknowledgment}  This work was supported in part by the Israel
Ministry of Absorption. 

\begin{table}
   	\begin{tabular}{ccccc}
		 & Solution I\tablenotemark[1] & Solution II\tablenotemark[2] 
		 & From Ref.\protect\cite{bramon95} 
        & From Ref.\protect\cite{veno98} \\
		\hline
		$c_W$ & $-0.19\pm0.04$ & $-0.19\pm 0.03 $& $-0.2 \pm 0.06$& \\
		$F_K/F_\pi$ & $1.16\pm 0.11$ & $1.16\pm 0.05$ & $1.22\pm 0.02$ & 
		$1.38\pm 0.22$  \\
		$F_\eta/ F_\pi$ & $1.09\pm 0.07$ & $1.14\pm 0.04 $& $1.06\pm 0.08$ & 
		 \\
		$F_0/F_\pi$ & $1.03\pm 0.06$ &$1.11\pm 0.04 $&  & $1.06 \pm 0.03$  \\
		$\theta_P$& $-(15.\pm 2.4)^o$ & $-(15.4\pm 1.8)^o$ & $ - 19.5^o$ & 
		$-(22\pm 3.3)^o$ \\
	\end{tabular}
	\caption{Values of parameters.}
	\tablenotemark[1]
	 {From the ratios $\Gamma(\eta\gamma\gamma)/ \Gamma(\phi\eta\gamma)$, 
   $\Gamma(\eta\gamma\gamma)/\Gamma(\pi^0\gamma\gamma)$, and
   $\Gamma(\eta'\gamma\gamma)/\Gamma(\eta\gamma\gamma)$. }\\ 
     \tablenotemark[2]{From a global fit of the decay widths of 
  the processes listed in Table II.}
   \protect\label{fit}
\end{table} 
\begin{table}
		\begin{tabular}{cccc}
		Decay &  $\Gamma_{exp}$(KeV) & $\Gamma_{calc}$(KeV) & 
		$\Gamma_{calc}$(KeV)  \\
		&    & solution I &  solution II\\
		    \hline
		$\rho\rightarrow \pi\gamma$ & $76\pm10$	& $76\pm 12$ & $76\pm 12$ \\
    	$\omega\rightarrow \pi\gamma$ & $716\pm43$ 	& *~$716\pm43$ & *~$716\pm43$  \\
		$\rho\rightarrow \eta\gamma$ & $58\pm11$ & $47\pm 11$ & $42\pm 10$  \\
		 $\omega\rightarrow \eta\gamma$ & $7.0\pm1.8$ & $6.\pm 1.6$ &
		  $5.5\pm 1.5$  \\
		$\phi\rightarrow \eta \gamma$ & $56.7\pm2.8$ & $61.4\pm 3$ &
		 $56.2 \pm 2.8$ \\
		$\phi\rightarrow \eta'\gamma$ & $\dagger\quad 0.54 \pm 
 		\begin{array}{c}
 			+0.29  \\
 			-0.23
 		\end{array}$ & $0.5\pm 0.14$ & $0.44\pm 0.14$  \\
		$\eta'\rightarrow \rho\gamma$ & $60.7 \pm 7.4$ & $74\pm 12$ & 
		$62\pm 9.1 $ \\
		$\eta'\rightarrow \omega\gamma$ & $6.07 \pm 0.74$ & $6.8\pm 0.8$ & 
		$6.0\pm 0.8$  \\
		$\pi^0\rightarrow \gamma\gamma$ & $7.8 \pm 0.55$ &*~$7.8 \pm 0.55$ & 
		*~$7.8 \pm 0.55$  \\ 
		$\eta\rightarrow \gamma\gamma$ & $460 \pm 40$  & $550\pm 70$ & 
		$490\pm 65$ \\
		$\eta'\rightarrow \gamma\gamma$ & $4290 \pm 190$ & $4430 \pm 280$  
		& $4050 \pm 260$  \\
		$K^{*0}\rightarrow K^0\gamma$ & $117\pm 10$ & $117\pm 10$ & 
		$117\pm 10$  \\
		$K^{*\pm}\rightarrow K^\pm\gamma$ & $ 50 \pm 5$ & $ 50 \pm 5$
		 & $ 50 \pm 5$  \\
		\end{tabular}¥
		\caption{Calculated decay widths with the parameter set
        solution I and solution II of Table \protect\ref{fit}. 
		Widths marked with asterisk were used to fix $g_P$ and $g_V$.
        Data are taken from \protect\cite{pdg98}. A value from the KLOE
        collaboration\protect\cite{pc},
        $\Gamma_{exp} (\phi\rightarrow \eta ' \gamma) = 
        (0.36 \pm 0.12)$ KeV yield practically the same results.}
 	\protect\label{vpgpred}
 \end{table}
 
 \section{Appendix}
The symmetry breaking terms ${\bar L}_A$ and ${\bar L}_V$ can be
constructed by using the nonet (rather than the octet) covariant
${\Delta}_\mu$. We demonstrate that for $L_A$. We write,
\begin{eqnarray}
       &  &\bar{L}_A =
           \nonumber\\  
       && W_1(X)\left( c_A Tr (\{B, \Delta_\mu\} \Delta^\mu) +
          d_A Tr (B \Delta_\mu B\Delta^\mu)\right) +
          \nonumber\\
       &&W_4(X) \left(c_A Tr (B \Delta_\mu) Tr (\Delta^\mu) +
          d_A Tr (B \Delta_\mu) Tr (B\Delta^\mu)\right)~,
          \label{labreak}
\end{eqnarray} 
From this expression the contributions of the $\eta_0$ and $\eta_8$
to the kinetic term is, 
   \begin{eqnarray} 
     L_{kin}^{08} =
         &&\kappa_{88}(\partial_\mu\eta_8)^2 +
\kappa_{00}(\partial_\mu\eta_0)^2 +
\kappa_{80}\partial_\mu\eta_8\partial^\mu\eta_0 +
         \nonumber \\
         && m^2_{88}\eta^2_8 + m^2_{00}\eta^2_0 + 2m^2_{80}\eta_8\eta_0,
                \label{}
         \end{eqnarray}
where the matrix $\kappa$ depends on the parameters $c_A$
and $d_A$. This expression gives rise to twofold $\eta -\eta '$ mixing,
one from the kinetic term and one from the nondiagonal mass matrix. We
first diagonalize the matrix $\kappa$ using the unitary transformation
         \begin{equation}
         \left(\begin{array}{c}
                        \eta_8  \\
                        \eta_0
                \end{array}\right) =
                \left(\begin{array}{cc}
                        \cos\lambda & \sin\lambda  \\
                        -\sin\lambda & \cos\lambda
                \end{array}\right) \left(
                \begin{array}{c}
                        \bar{\eta}_8  \\
                        \bar{\eta}_0
                \end{array}\right) = \Upsilon\left(\begin{array}{c}
                        \bar{\eta}_8  \\
                        \bar{\eta}_0
                \end{array}\right)
         \end{equation}  
This leads to,
\begin{equation}
    L_{kin}^{08}  =
                \kappa_8(\partial_\mu\bar{\eta}_8)^2 + \kappa_0
                (\partial_\mu\bar{\eta}_0)^2  +  (\bar{\eta}_8,
\bar{\eta}_0)
                \Upsilon^{-1}{\cal M}^2\Upsilon
                \left(\begin{array}{c}
                        \bar{\eta}_8  \\
                        \bar{\eta}_0
                \end{array}\right)
                \label{}
         \end{equation}
where $\kappa_i$ are the eigenvalues of the matrix $\kappa$. Now to
restore the standard normalization of the kinetic term we rescale the
pseudoscalar fields
\begin{equation}
        \pi\Rightarrow z_\pi\pi,\qquad K\Rightarrow z_s K,\qquad
\bar{\eta}_8
        \Rightarrow z\hat{\eta}_8~,\qquad \bar{\eta}_0 \Rightarrow
f\hat{\eta}_0
        \label{rescal}
\end{equation}
where $z=1/\sqrt{\kappa_8}, f=1/\sqrt{\kappa_0}$. In other words the 
fields
$\bar{\eta}_i$ and $\hat{\eta}_i$ are related by nonunitary
transformation (matrix) $R = diag(z,f)$. Therefore, the $\hat{\eta}$ 
mass matrix has the (nondiagonal) form
\begin{equation}   
        \hat{{\cal M}}^2 = R\Upsilon^{-1}{\cal M}^2\Upsilon R
        \label{}
\end{equation}
and the Goldstone field kinetic term now reads $(1/2)[(\partial_\mu
\pi)^2 + (\partial_\mu K)^2 +(\partial_\mu \hat{\eta}_8)^2 +
(\partial_\mu\hat{\eta}_0)^2]$. As a last step we relate the
$\hat{\eta}_i$ to the physical fields $\eta$ and $\eta'$ which are
eigenvectors of the mass matrix $\hat{{\cal M}}^2$,
\begin{equation}
        \left(  
        \begin{array}{c}
                \hat{\eta}_8  \\
                \hat{\eta}_0
        \end{array}
        \right) =\Omega \left(
        \begin{array}{c} 
                \eta  \\
                \eta '
        \end{array}
        \right) ~,
        \label{}   
\end{equation}
where,
\begin{equation}
        \Omega = \left(
        \begin{array}{cc}
                \cos\chi & \sin\chi  \\
                -\sin\chi & \cos\chi
        \end{array}    
        \right)~.
        \label{}
\end{equation}  
The relations between the $\eta_8$ and $\eta_0$
and physical fields $\eta,~ \eta'$ is then given by,
\begin{equation}
        \left(  
        \begin{array}{c}
                \eta_8  \\
                \eta_0
        \end{array}
        \right) = \Theta\left(
        \begin{array}{c} 
                \eta  \\
                \eta'
        \end{array}
        \right)~.
        \label{}   
\end{equation}
where,
\begin{equation}
        \Theta = \left(
        \begin{array}{cc}
    z\cos\lambda\cos\chi-f\sin\lambda\sin\chi &
      z\cos\lambda\sin\chi+f\sin\lambda\cos\chi  \\
         -z\sin\lambda\cos\chi-f\cos\lambda\sin\chi &
         -z\sin\lambda\sin\chi+f\cos\lambda\cos\chi
        \end{array}
        \right)~.
        \label{tetatr}
\end{equation} 
Clearly, the transformation $\Theta$ is nonunitary and can not be
written in the so called two angle form of Refs.\cite{bramon97,escrib99}
since $\Theta^2_{i1} + \Theta^2_{i2} \neq 1$.

The pseudoscalar meson matrix has the form of Eqn.\ref{pnonet}
but with $X$ and $Y$ defined as, 
\begin{eqnarray} 
        && X_\eta = z\cos\lambda\cos\chi-f\sin\lambda\sin\chi +
        \sqrt{2}r(-z\sin\lambda\cos\chi-f\cos\lambda\sin\chi)~,
     \nonumber  \\
    &&X_{\eta'} = z\cos\lambda\sin\chi+f\sin\lambda\cos\chi  +
    \sqrt{2}r(-z\sin\lambda\sin\chi+f\cos\lambda\cos\chi)~,
        \nonumber \\
    &&Y_\eta = -2(z\cos\lambda\cos\chi-f\sin\lambda\sin\chi) +
    \sqrt{2}r(-z\sin\lambda\cos\chi-f\cos\lambda\sin\chi)~,   
        \nonumber  \\
    &&Y_{\eta'} =
    -2(z\cos\lambda\sin\chi+f\sin\lambda\cos\chi) +
   \sqrt{2}r(-z\sin\lambda\sin\chi+f\cos\lambda\cos\chi)~.
        \label{nxy}
\end{eqnarray}
Clearly, for $\lambda = 0$ and $f = 1$ the expressions
\ref{nxy} reduce to the ones in Eqns. \ref{xy}.

 	\end{document}